\documentclass[ aps prd, showpacs, superscriptaddress, twocolumn ,groupedaddress]{revtex4-1} 
\usepackage{graphicx}	
\usepackage{multirow}
\usepackage{amssymb}
\usepackage{dcolumn}
\usepackage{color}
\usepackage{subfigure, rotating, bm, array}
\usepackage[pagebackref=false, colorlinks=true]{hyperref}
\hypersetup{
linkcolor=blue,     
citecolor=blue,     
urlcolor=blue}      
\def\beq{\begin{equation}}
\def\eeq{\end{equation}}
\def\bea{\begin{eqnarray}}
\def\eea{\begin{eqnarray}}

\begin{document}
\title{Gravitational Collapse of Massless Vector Field with Positive Cosmological Constant}

\author{Tapobroto Bhanja}
\email{tapobroto.bhanja@gmail.com}
\affiliation{International Center for Cosmology, Charotar University of Science and Technology, Anand, Gujarat - 388421, India}


\author{Ameya Kolhatkar}
\email{kolhatkarameya1996@gmail.com }
\affiliation{PDPIAS, Charotar University of Science and Technology, Anand, Gujarat - 388421, India}


\date{\today}

\begin{abstract}
\noindent
We investigate the dynamics of homogeneous gravitational collapse of a massless vector field in the presence of a positive cosmological constant $\Lambda$. 
The corresponding density function $\rho (a)$ obtained for the massless vector field is inversely proportional to the fourth power of the scale factor $a (t)$. 
The variation of the scale factor shows that for $0\, \le\Lambda < 1$, we obtain the gravitational collapse of the vector fields leading singularity formation 
in a {\it finite} comoving time resulting in a {\it Blackhole} such that with increasing $\Lambda$, the singularity formation time, $t_s$ increases. For  
$\Lambda = 1$, we obtain $a(t) = 0$, thus limiting the maximum value of $\Lambda$, (w.r.t the initial density $\rho_0$) for which the system could collapse under 
gravity.  \\
\\\begin{large}
\end{large}
 {\bf key words}: Gravitational Collapse, Massless Vector fields, Blackhole 
\end{abstract}

\maketitle
\section{Introduction}
\noindent

Any massive body, having a mass greater than a particular limit, collapsing under the influence of its own gravity leads to the formation of singularity. 
The weak and strong cosmic censorship conjectures say that the singularity thus formed would be hidden from a distant observer by the event horizon 
formed during/after the formation of singularity during the gravitational collapse and no null geodesic can escape  the singularity
and reach an asymptotic/distant observer [1-4]. However, it is theoretically possible that there exists certain cases where null geodesics can 
escape the singularity and reach a distant observer where the collapsing system has a certain minimum amount of radial inhomogeneity in density. 
The visible singularity thus formed,  could either be locally or globally visible depending upon the formation time of the trapped surfaces. For the 
singularities to be {\it globally} visible, the null geodesics must come out from the singularity and reach a distant observer before it crosses the
apparent horizon or the boundary of the trapped surfaces. To put it in a simpler way, the trapped surfaces should be formed at infinite comoving time. 
It may happen that the null geodesics escape the singularity and before reaching a distant observer, it encounters the trapped surface and fall back in
towards the singularity. The singularity would be {\it locally} visible to any observer who lies within the apparent horizon. However, these locally visible
singularities are not astrophysicaly much relevant, while for globally visible ones, the outgoing  null  geodesics can in principle reach an asymptotic observer 
and thus have a much greater relevance in the realm of astrophysics.

The context of visibility of singularities formed by gravitational collapse has gained much interest and importance in recent times [5-17].  
It has been shown that  dust clouds  having a minimum or sufficient inhomogeneity collapsing spherically can result in the formation of a 
Naked Singularity (NS)  where null geodesics can escape the trapped surfaces and reach distant observers, thus being visible to them [18-19]
The prime reason being that if these spacetime singularities are visible to a distant observer, the visible null geodesics would bear
the possible physical and detectable signatures of quantum effects (of gravity) which are expected to have strong impacts in the region of ultra-strong 
gravitational fields and thus providing us with the much needed insights on the physical effects of quantum gravity on a spacetime manifold.  
The importance of study of quantum effects of gravity and correspondingly the visibility of singularities could be well argued form the 
interpretation of the singularity theorem which predicts the breakdown of general relativity (GR) at singularities formed during gravitational 
collapse (or otherwise) where the physical variables (like matter  or radiation density) blows up, rendering the system unphysical. 
Thus, formation and existence of spacetime singularities puts a limit/boundary on the validity of GR itself in regions of high gravity and curvature 
of the spacetime manifold. One could interpret the blowing up of physical quantities as particle productions in higher curvature regions and thus
bringing in the quantum effects into the game, inevitably. It could be further argued that study the laws of physics in such high curvature and gravity regions 
in the context of the Standard Model of particle physics (SM) and the interactions therein, would bring further insights into the dynamics
of the observable/physical variables at the extreme vicinity of  the spacetime singularities.

In this present endeavour we study a toy model of homogenous gravitational collapse of a massless vector field in the presence of a positive 
cosmological constant $\Lambda$.  The toy model satisfies both the weak and strong energy 
conditions required for gravitational collapse. The 
positive cosmological constant plays an important role in the nature of the singularity formed. For $0 \le \Lambda < 1$ this model satifsies the
the cosmic censorship conjecture and results in the formation of a BH. Our work is organised as follows: In Sec. II, we deal with the basic dynamical equations governing the collapse 
on FLRW spacetime background. We show explicitly that the energy density falls off as fourth power of the scale factor $a (t)$. Sec III is dedicated 
to the study of the causal structures of the singularity. In Sec. IV we summarise our work with concluding remaks.  Throughout the paper we 
have used the geometrised units, viz. $8 \pi G = c = 1$.



\section{Dynamics of Collapse}\label{sectiontwo}
 \,

We begin with the Lagrangian density for the massless  a vector field $(A_\mu)$, 
\begin{equation}
\label{1}
 {\cal L_M}  \,=  \,- \, \frac{1}{4}\, F_{\mu\nu}\,F^{\mu\nu},
\end{equation}
where $F_{\mu\nu}$ is the electromagnetic field strength tensor and $F_{\mu\nu}\,F^{\mu\nu}$ denotes its kinetic term and  
study the gravitational  collapse of the vector field $A^\mu$ and the  spacetime background governed by the spatially flat
Friedmann–Lemaitre–Robertson–Walker  (FLRW) metric given by,
\begin{equation}
\label{2}
ds^2 \, = \, -  \,\,dt^2 + a^2 \,dr^2 + R^2  \,d{\Omega}^2,
\end{equation}
where $R^2 d{\Omega}^2 = a^2 r^2 (d{\theta}^2 + sin^2 \theta \, d{\phi}^2$). Here $a \equiv a(t)$ is the scale 
factor with the boundary conditions $a(0) = 1$ and $a(t_s) = 0$,  $t_s$ being the singularity formation time. 
$R \equiv R(r, t)$ is the $physical$ radius and $r$ is the $comoving$ radius 
related by the relation: $R(r, t) = r \,a (t)$. The energy-momentum tensor can be calculated from the Lagrangian
density for the vector field is given by
\begin{equation}
\label{3}
T_{\mu\nu}  \,=  \,- \, \frac{2}{\sqrt{- g}} \, \frac{\delta (\sqrt{- g}\, \cal L_M)}{\delta g^{\mu\nu}},
\end{equation}
such that the the components of the energy momentum tensor is given by $T_{\mu\nu} =$ diag $\, (\rho, \, p,\, p,\, p)$, where
$\rho$ and $p$ are the density and pressure, respectively, and in presence of cosmological constant ($\Lambda$) are subsequently expressed by the relations
\begin{eqnarray}
\label{4}
&&\rho \, =  \, \frac{3}{2}\Big[a^2  \dot {A}^2 + 2 \, a \, \dot{a}\, A \, \dot {A} \Big]  \,  =  \, \frac{3 \dot{a}^2}{a^2} + \Lambda, \nonumber\\
&&p \, =  \,\frac{1}{2} \Big[a^2  \dot {A}^2 + 2 \, a \, \dot{a}\, A \, \dot {A} \Big] \,  =  \,- \,\frac{2\ddot{a}}{a} - \frac{\dot{a}^2}{a^2} -  \Lambda,
\end{eqnarray}
where the overdot represents the time derivative of the functions w.r.t. time. It can be easily checked that the massless vector field
follows the expected equation of state for radiation, viz, $p/ \rho = \omega = 1/3$. The Klein Gordon (KG) equation (equations
of motion)  is given by: 
\begin{equation}
\label{5}
\partial_\mu \big(\sqrt{- g} \, F^{\mu\nu}\big)  \,=  \,0,
\end{equation}
where $\sqrt{- g} = [a (t)]^3$ for FLRW background. In case of FLRW metric, it can be seen that there are no dynamics of the `scalar potential' $\phi$.  
Thus, the above equation (\ref{5}) now transforms to
\begin{equation}
\label{6}
\ddot {A}  \,+  \,3  \,H \, \dot {A}  \,=  \,0,
\end{equation}
where $H = (\dot{a}/a)$ is the Hubble constant and we have considered the spatial components of the vector field $(A_i)$ is a function of the scale factor $a$ and time $t$ only
and has the same value A (i.e., $A_i = A$ and $\frac{\partial A}{\partial r} = 0$, $r$ being the comoving radius). The above 
equation (\ref{6}) can also be derived from the Einstein's field equations (\ref{4}) such that  $\dot A \ne 0$, identically. Thus, it is 
viable to observe/infer that the KG equation is not a free equation and can be deduced from the expressions of the energy density ($\rho$) and pressure 
$(p)$.   Using the chain rule: $\dot {A} = A' \dot {a}$, and both the equations from (\ref{4}), we get: 
\begin{equation}
\label{7}
\rho  \, +  \,p  \,=  \,2 \, \Big[a^2  \dot {A}^2 + 2 \, a \, \dot{a}\, A \, \dot {A} \Big].
\end{equation}
We can obtain the dynamics of the collapse from the first equation in (\ref{4}) as, 
\begin{equation}
\label{8}
\dot {a}  \,=  \,- \, a \, \sqrt {\frac{\rho \, (a) - \Lambda}{3}}, 
\end{equation}
and thus correspondingly  by differentiating the above equation we obtain:
\begin{equation}
\label{9}
\ddot {a} \, = \, \frac{a}{3} \, \big(\rho - \Lambda + \frac{a}{2}\,{\rho}'\big).
\end{equation}
Further, we get from the above equation (\ref{4}), we get the following relation, 
\begin{equation}
\label{10}
\rho \, -  \,p  \,=  \,\Big[a^2  \dot {A}^2 + 2 \, a \, \dot{a}\, A \, \dot {A} \Big],
\end{equation}
and using Eqn. (\ref{8}) in Eqn. (\ref{7}), we obtain: 
\begin{eqnarray}
\label{11}
\rho\, \Big[1 -  \frac{2}{3} \, (a\, A')^2 &-& \frac{4}{3}\, a^3 A A' \Big]  \,+ \, p   \nonumber\\ &=&   \, \Big[\frac{2}{3} \, (a\, A')^2 + \frac{4}{3} a^3 A A' \Big]\Lambda.
\end{eqnarray}
Using the equation (\ref{4}) we get: 
\begin{equation}
\label{12}
p = - \rho -  \, \frac{a {\rho}'}{3}
\end{equation}
With a little algebra involving  equations  (\ref{9}), (\ref{10}), (\ref{11}) and (\ref{12}) we obtain, 
\begin{equation}
\label{13}
\frac{{\rho}'}{{\rho}}\,= - \frac{4}{a},
\end{equation}
which immediately gives us,
\begin{equation}
\label{14}
\rho \, (a)   \,=  \, \frac{{\rho}_0}{[a (t)]^4},
\end{equation}
where ${\rho}_0$  is an integration constant (${\rho} = {\rho}_0$ at $t = 0$). Consequently, we get
\begin{equation}
\label{15}
\dot{a} = - a \sqrt{\frac{ {\rho}_0}{3\,  a^4} - \frac{\Lambda}{3}},
\end{equation}
which when integrated leads to the explicit expression of the scale factor $a (t)$. However, 
it is to be noted that the above differential equation cannot be solved analytically and hence,
we go for numerical analysis to get the variation of the scale factor $a (t)$ and  correspondingly the nature of the singularity. 
Thus, we find that for the massless vector fields $A^\mu$ in
presence of a positive cosmological constant $\Lambda$,
\begin{eqnarray}
\label{16}
\rho \, (a) \propto  \, a^{- 4},
\end{eqnarray}
showing that during collapse as $a (t)$ decreases, $\rho  \,(a)$ diverges and ultimately 
blows up as $a (t) \rightarrow 0$.

\section{Causal Structure of Singularity}\label{sectionfour}

The dynamics of the scale factor $a(t)$ and that of the energy density  $\rho(a)$ for the  massless vector fields in 
presence of the cosmological constant $\Lambda$ shows that as time $t$ increases from $0$, $a(t)$ decreases and energy density 
$\rho \, (a)$ increases. As $a (t) \rightarrow 0$ we approach singularity and the nature of this strong gravity region can be explored 
by investigating formation of trapped surfaces around the singularity. At any instant of time $t$, trapped surfaces are not formed when
\begin{equation}
\label{17}
\frac{\rho \, R^2}{3} = \frac{r^2 \,a^2}{3}\,  \Big(\frac{{\rho}_0}{a^4}\Big)  < 1,
\end{equation}
where we have considered $\rho_0 = 1$ for simplicity and $R$ is the physical radius of the collapsing cloud. 
The expansion scalar $\theta_l$ for the outgoing null geodesic congruence for the case of 
a massless vector fields is then given by, 
\begin{eqnarray}
\label{18}
\theta_l = \frac{2}{R} \, \Big(1 - \sqrt{\frac{\rho}{3}}\, R\Big), 
\end{eqnarray}
with equation  (\ref{16}) implying the positivity of the above equation  (\ref{17}). The expansion scalar is positive (i.e., $\theta_l > 0$) when $a(t)$, if 
it satisfies the following equation, 
\begin{eqnarray}
{r_c}^2 < \lim_{a \to 0} \frac{3}{a^2\, \rho}, 
\end{eqnarray}
where ${r_c}$ is defined as the largest comoving radius of the collapsing system. 
Hence we can observe that there is always a causal connection when we have positive expansion 
scalar. The nature of singularity formed is obtained by substituting the value of $\rho$ from Eqn.  (\ref{14}) in Eqn.  (\ref{8})
and then solving the corresponding differential equation, thus obtained with the boundary condition: $a (t) = 1, t = 0$. The variation 
of $a (t)$ is plotted against $t$ for five different values of $\Lambda$ viz, 0, 0.2, 0.4, 0.8 \& 1: 
\begin{figure}[htbp]
\centerline{\includegraphics[scale=.5]{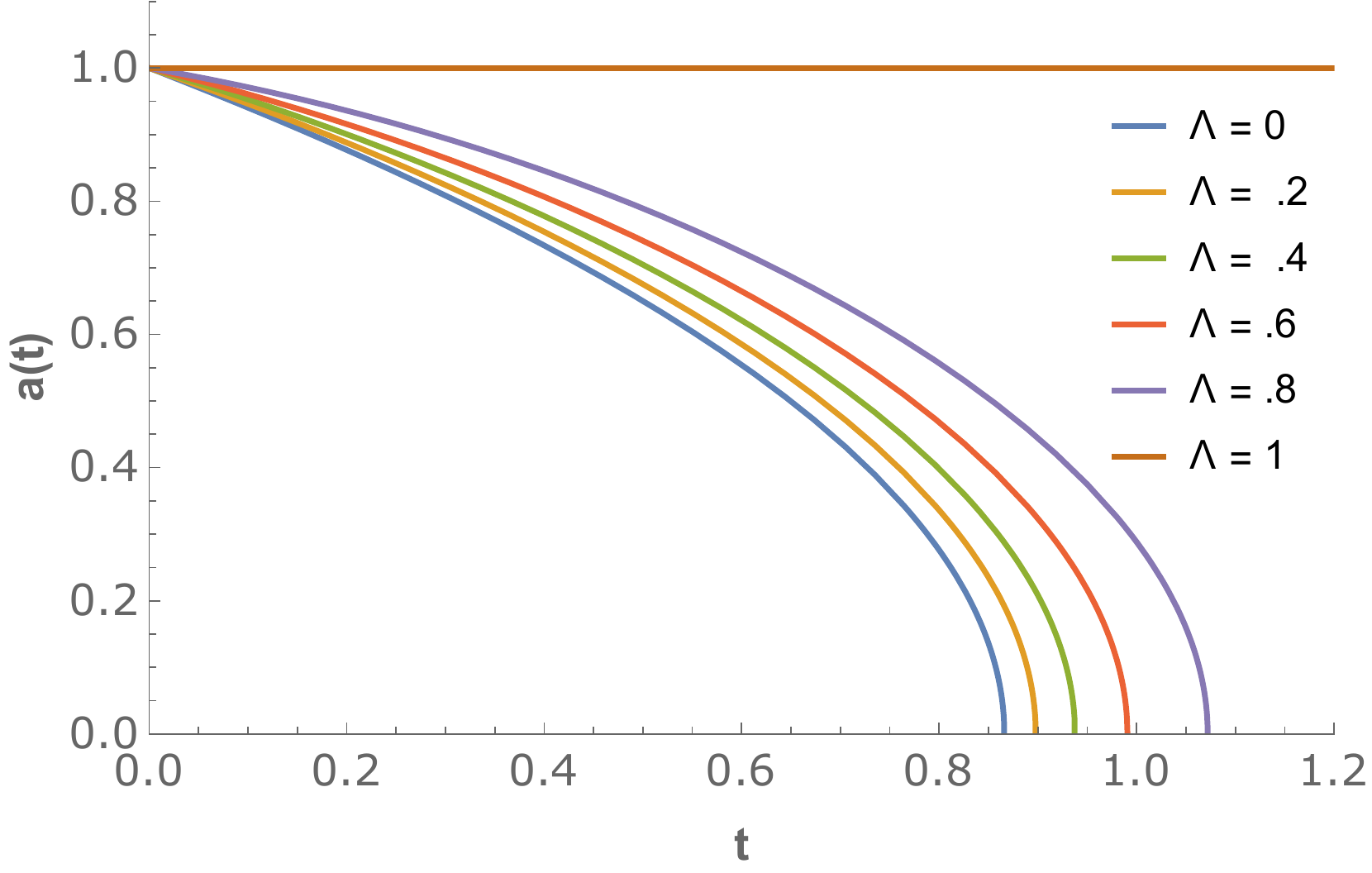}}
\caption{Variation of scale factor with time for different $\Lambda$}
\label{fig1}
\end{figure}

It is interesting to note that, for $\Lambda = 0, 0.2, 0.4, \& \, 0.8 $, $a (t)$ becomes zero at a {\it finite} comoving  time. For 
$\Lambda = 1$,  no singularity is formed in time. The variations of F/R with time $t$, for the same values of $\Lambda$ are  plotted as:
\begin{figure}[htbp]
\centering
{\includegraphics[scale=.45]{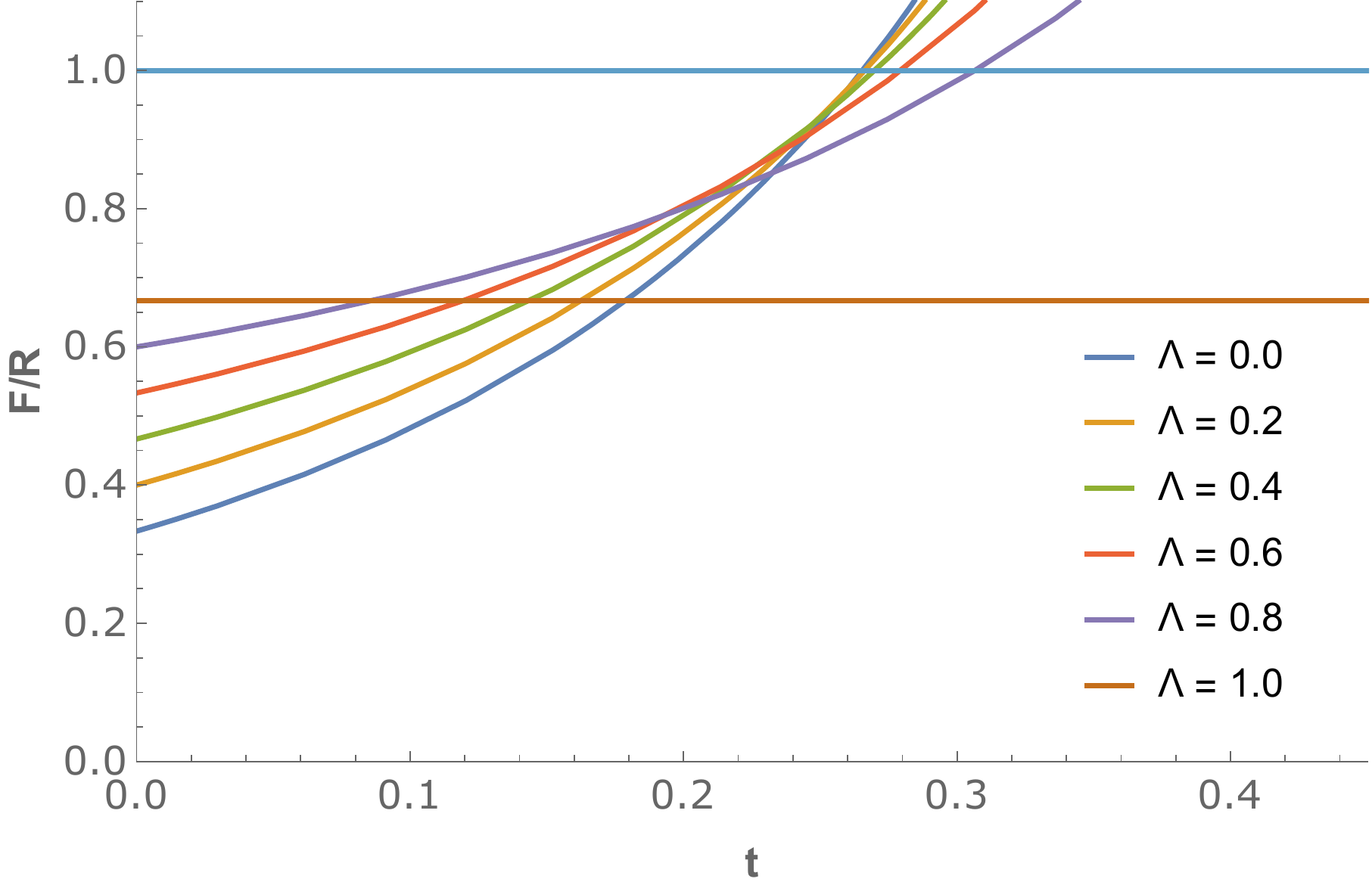}}
\caption{Variation of F/R with time (t) for different time and  $\Lambda$ at r = 1}
\label{fig2}
\end{figure}


\section{Conclusion}

We study the toy model of a homogenous gravitational collapse for a massless vector field $A^\mu$ in presence of cosmological constant $\Lambda$,
where the spatially flat FLRW metric seeded by the massless vector field  $A^\mu$   and the cosmological constant $\Lambda$.
 From the Euler Lagrange (EL) equation of motion (\ref{5}), we get that there are no dynamics of the scalar potential. 
Since we are considering the homogenous collapse, we do not have any space dependence on the vector field $A^\mu$ and it is only a function of time $t$. 
Our present endeavour provide us with some interesting insights.
 First, we obtained the density function in an algebraic form $\rho \, (a) \propto  \, a^{- 4}$, and by solving the 
corresponding differential equation numerically, we obtain the corresponding variation of the scale factor $a (t)$ w.r.t $t$ as shown in Fig. (\ref{fig1}).  
It is observed that as we increase the value of $\Lambda$, the singularity formation time increases, which depicts that $\Lambda$ acts as a 
repulsive force, in general, and thus preventing gravitational collapse of the vector fields. As the singularities are formed in finite comoving time, the gravtitational 
collapse for the massless vector fields would always lead to the formation of Blackholes. It is to be noted here that the chosen values or range of $\Lambda$
depends upon the value of the integration constant $\rho_0$ (which represents the initial density of the collapsing system) appearing in Eqn. (14). We 
have chosen $\rho_0 = 1$ throughout our calculations. 
 If the chosen value of $\rho_0$ is greater than or less than 1, then the range of  $\Lambda$ and the value of the respective 
singularity formation time $t_s$ would change correspondingly, however the nature of the variation of $a (t)$ with $t$ would remain the same. 
If we consider the value of $\Lambda$ more than the value of the initial energy density then we would not have any collapsing scenario. 
Further, we do not get any physical situation for $\Lambda > \rho_0$, as then we would encounter complex quantities, which are not considered 
physical (as evident from equation (\ref{8}) and we do not observe any gravitational collapse.

Second, the presence of the cosmological constant $\Lambda$ shows that as it  increases the singularity formation time $t_s$ increases. 
As the scale factor falls with time, the plot of $F/R$ vs. time $t$  (Fig. (\ref{fig2})) shows that as time increases, the apparent horizon starts moving inwards (towards $R = 0$), 
and it crosses  each of the comoving radius (except the comoving radius corresponding to $r = 0$) before the formation of singularity. 
Hence, there are permanent trapped surfaces formed from which no null geodesic can escapeafter the apparent horizon crosses the comoving radius  
leading to formation of blackholes. This is a general feature of the situation for $\Lambda < 1$.  

When the value of $\Lambda$ equals the value of the initial density $\rho_0$, then we get the straight line as seen 
in Fig. (\ref{fig1}), meaning that the scale factor does not change (increase or decrease) with time and we get a static universe. 
Further, for $\Lambda = 1$, we observe that ``F/R" does not
change with time and has a fixed value around 0.63. This corresponds to the case where the presence of the 
cosmological constant prevents the collapse and hence a singularity. We obtain a ball of radiation, forming a stable equilibrium state due to the two opposite forces at work. 
As for this case, the value of ``F/R" is less than 1, this (stable) ball of radiation would be visible to a distant observer.

\end{document}